\documentclass[twocolumn,showpacs,preprintnumbers,prb]{revtex4}
\usepackage{graphicx,bm}

\def\gz{\ifmmode{Z\hskip -4.8pt Z}
  \else{\hbox{$Z\hskip -4.8pt Z$}}\fi}

\newcommand{\be}{\begin{equation}}
\newcommand{\ee}{\end{equation}}
\newcommand{\bea}{\begin{eqnarray}}
\newcommand{\eea}{\end{eqnarray}}

\begin{document}

\title{Features of spin-charge separation in the equilibrium conductance
through finite rings}
\author{Juli\'{a}n~Rinc\'{o}n, A.~A.~Aligia and K.~Hallberg}
\affiliation{Centro At\'{o}mico Bariloche and Instituto Balseiro, Comisi\'{o}n Nacional
de Energ\'{\i}a At\'{o}mica, 8400 Bariloche, Argentina}
\date{\today}

\begin{abstract}
We calculate the conductance through rings with few sites $L$ described by
the $t-J$ model, threaded by a magnetic flux $\Phi$ and weakly coupled to
conducting leads at two arbitrary sites. The model can describe a circular
array of quantum dots with large charging energy $U$ in comparison with the
nearest-neighbor hopping $t$. We determine analytically the particular
values of $\Phi $ for which a depression of the transmittance is expected as
a consequence of spin-charge separation. We show numerically that the
equilibrium conductance at zero temperature is depressed at those particular
values of $\Phi $ for most systems, in particular at half filling, which
might be easier to realize experimentally.
\end{abstract}

\maketitle


\section{Introduction}

Recent advances in nanotechnology allow the fabrication of different
nanostructures, motivated either by technological interest or the
possibility of testing theories for strongly correlated electrons. One
example is the realization of the Kondo effect in systems with one quantum
dot (QD).~\cite{gold1,cro,gold2,wiel} Another one is a study of the
metal-insulator transition in a chain of 15 QD's.~\cite{kou} Systems of a
few QD's have been proposed theoretically as realizations of the two-channel
Kondo model,~\cite{oreg,zit} the so called ionic Hubbard model,~\cite{ihm}
and the double exchange mechanism.~\cite{mart} Recently it has been studied
how the Kondo resonance splits in two in a system of two QD's, one of them
non-interacting.~\cite{dias,vau}

It is known that strong correlation effects invalidate the conventional
quasiparticle description of Fermi liquids and in one dimension lead to a
fractionalization of the electronic excitations into pure charge and spin
modes, as it has been shown using bosonization.~\cite%
{Haldane,Schulz,kv,gia,kurt} In this field theory, it becomes clear that in
general (except in some particular models~\cite{rc}) the Hamiltonian at low
energies can be separated into independent charge and spin parts. While this
separation is an asymptotic low-energy property in an infinite chain, exact
Bethe ansatz results for the Hubbard model in the limit of infinite Coulomb
repulsion $U$ have shown that the wave function factorizes into a charge
part and a spin part for any size of the system.~\cite{ogata}

Several experiments were reported that find indirect evidences of
spin-charge separation~\cite{Voit1,Voit2,Kim,Taillefer,Qimiao}, and it could
also be potentially observed~\cite{kurt} in systems such as cuprate chains
and ladder compounds,~\cite{DagottoRice} and carbon nanotubes.~\cite{egger}

From the theoretical point of view, several calculations involving rings
have been made. The real-time evolution of electronic wave packets in
Hubbard rings has shown a splitting in the dispersion of the spin and charge
densities as a consequence of the different charge and spin velocities.~\cite%
{Jagla1,uli} Pseudospin-charge separation has also been studied theoretically in quasi-one-dimensional quantum gases of fermionic atoms.~\cite{recati,ke}

The transmittance through Aharonov-Bohm rings of length $L$ modeled by a
Tomonaga-Luttinger liquid and connected to conducting leads at $0$ and $L/2$
has been studied by analytical methods.\cite{Jagla,meden} It is found that
the transmittance integrated over an energy window shows dips when the
threaded magnetic flux $\Phi $ corresponds to particular fractional values
of the flux quantum $\Phi _{0}=hc/e$. This is rather striking because in the
non-interacting case, one has a dip in the transmittance only when the
applied flux $\Phi =\Phi _{0}/2$, for which the conductance vanishes due to
a negative interference of the waves traveling through both arms of the
interferometer. Jagla and Balseiro obtained that the values of the flux at
the dips are multiples of $\Phi _{0}v_{s}/v_{c}$, where $v_{c}$ ($v_{s}$) is
the charge (spin) velocity, when $v_{s}/v_{c}$ is a simple fraction. This
allows an appealing simple \textquotedblleft classical\textquotedblright\
interpretation of the phenomenon: the electrons enter the ring at position $0
$, splitting into charge and spin components, which travel independently
inside the ring, until they recombine at $L/2$ before leaving the ring. When
the difference between the Aharonov-Bohm phases captured by the charges
traveling in both possible senses of rotation is an odd multiple of $\pi $,
the transmittance is depressed. More recent work, however, indicates that
the relevant ratio for the dips is in general $v_{s}/v_{J}$, where $v_{J}$
is the current velocity.~\cite{meden}

Numerical calculations of the transmittance through finite rings described
by the $t-J$ model, integrated over an energy window also show clear dips
for fractional applied fluxes.~\cite{nos1} This model is expected to be a
realistic description for a ring of quantum dots if the charging energy $U$
is large in comparison with the nearest-neighbor hopping $t$. Recently we
have discussed the extension of these results to ladders of two legs as a
first step to higher dimensions.~\cite{julian}

All the above calculations used a formalism valid at equilibrium and zero
temperature, and an integration over a finite energy window to obtain the
dips. In particular in Ref.~\onlinecite{nos1} this window contained all
low-energy spin excitations with the same charge quantum numbers $n_{l}$
(see Section III)). In principle, this integration can be justified invoking
a finite voltage bias or temperature. However, in the interacting case,
under an applied bias, it is not clear that the total current through the
device can be obtained integrating the \emph{equilibrium} transmittance. In
particular, it might happen that a particle injected to the ring leaves it
in an excited spin state after leaving it. This process is not taken into
account in the calculations. The effect of temperatures of the order of the
spin excitation energy is also difficult to predict. These shortcomings
raise the question of whether the dips can be really observed in an
experimental setup.

 In this work, we analyze the origin of the dips in the
transmittance as a function of applied flux in finite rings described by a
strongly correlated model and in particular, the $t-J$ model. We discuss the
conditions for which the intensity of the first peak in the equilibrium
conductance at zero temperature as a function of the gate voltage has a flux
dependence with characteristic dips as a consequence of spin-charge
separation.

In Section II we present the model, the formalism used to calculate the
conductance, and some general statements on the conditions for which dips or
reduced conductances are expected. In Section III we discuss the model with $%
J=0$ (equivalent to the Hubbard model with infinite $U$), for which definite
conclusions can be drawn on the basis of its exactly known energy spectrum.
Section IV contains numerical results for some particular systems at which
dips in the conductance at certain fluxes are expected at equilibrium and
small temperatures. Section V is a summary and discussion.

\section{Model and relevant equations}

\subsection{Hamiltonian}

\begin{figure}[tbp]
\includegraphics[width=8cm]{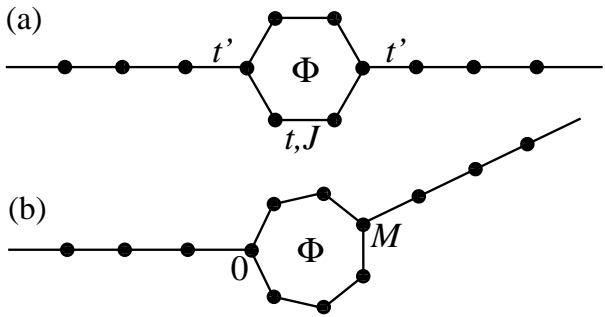}
\caption{Scheme of the systems studied numerically. (a) $L=6$, (b) $L=7$. In
both cases $M=3$.}
\label{dfhepta}
\end{figure}

We consider a ring of $L$ sites, weakly connected to non-interacting leads
at sites $0$ and $M$ (Fig.~\ref{dfhepta}). Usually $M=L/2$ was taken,~\cite%
{Jagla,meden,nos1} but we will see that $L$ odd and/or $M\neq L/2$ lead to
interesting new results.

The Hamiltonian can be written as 
\begin{equation}
H=H_{\mathrm{ring}}+H_{\mathrm{leads}}+H_{\mathrm{links}}.  \label{esh}
\end{equation}%
The first term describes the isolated ring, with on-site energy
modified by a gate voltage $V_{g}$, and hoppings modified by the phase $\exp
(i\phi /L)$ due to the circulation of the vector potential. For most of the
results of this paper we use the $t-J$ model to describe the ring, 
\begin{eqnarray}
H_{\mathrm{ring}} &=&-eV_{g}\sum_{i\sigma }c_{i\sigma }^{\dag }c_{i\sigma
}-t\left( c_{i+1\sigma }^{\dag }c_{i\sigma }e^{i\phi /L}+\mathrm{H.c.}%
\right)   \nonumber \\
&&+J\sum_{i}\left(\mathbf{S}_{i}\cdot \mathbf{S}_{i+1}-\frac{1}{4}\right).  
\label{ehr}
\end{eqnarray}%
where $\phi =2\pi \Phi /\Phi _{0}$, $\mathbf{S}_{i}$ is the spin operator at
site $i$ and double occupancy is not allowed at any site of the ring. The
second term corresponds to two tight-binding semi-infinite chains for the
left and right leads 
\begin{equation}
H_{\mathrm{leads}}\!=\!-t\!\!\!\sum_{i=-\infty ,\sigma
}^{-1}\!\!\!a_{i-1,\sigma }^{\dag }a_{i,\sigma }\!-t\!\!\!\sum_{i=1,\sigma
}^{\infty }\!a_{i,\sigma }^{\dag }a_{i+1,\sigma }\!+\!\mathrm{H.c..}
\label{ehl1}
\end{equation}

The third term in Eq.~(\ref{esh}) describes the coupling of the left (right) lead with
site 0 ($M$) of the ring 
\begin{equation}
H_{\mathrm{links}}=-t^{\prime }\sum_{\sigma }(a_{-1,\sigma }^{\dag
}c_{0\sigma }+a_{1,\sigma }^{\dag }c_{M\sigma }+\mathrm{H.c.}).  \label{ehl2}
\end{equation}

\subsection{Conductance}

To calculate the conductance through the ring, one needs, in principle, to
know some Green's functions of the complete system.~\cite{meir} However,
when the ground state of the isolated ring is non-degenerate, and the
coupling $t^{\prime }$ between the leads and ring is weak, the equilibrium
conductance at zero temperature can be expressed to second order in $%
t^{\prime }$ in terms of the retarded Green's function for the isolated ring
between sites $i$ and $j$: $G_{i,j}^{\mathrm{R}}(\omega )$.~\cite{Jagla,ihm}
For an incident particle with energy $\omega =-2t\cos k$ and momentum $\pm k$%
, the transmittance reads 
\begin{equation}
T(\omega ,V_{g},\phi )=\frac{4t^{2}\sin ^{2}k|{\tilde{t}}(\omega )|^{2}}{%
\left\vert [\omega -{\epsilon }(\omega )+te^{ik}]^{2}-|{\tilde{t}}%
^{2}(\omega )|\right\vert ^{2}},  \label{tra}
\end{equation}%
where 
\begin{equation}
\epsilon (\omega )=t^{\prime \,2}G_{00}^{\mathrm{R}}(\omega )\text{, }\tilde{%
t}(\omega )=t^{\prime \,2}G_{0M}^{\mathrm{R}}(\omega ),  \label{et}
\end{equation}%
play the role of a correction to the on-site energy at the extremes of the
leads and an effective hopping between them respectively.

Although derived from perturbation theory, this equation is in fact exact
for a non-interacting system. However, in general, for an odd number of
electrons, the ground state of the isolated ring is Kramers degenerate for a
system with time reversal symmetry and the equation ceases to be valid,
missing the physics of the ensuing (non-perturbative) Kondo effect in which
the electrons of the leads screen the spin of the ring.~\cite{ihm,lobos} 
Nevertheless, the characteristic energy of this Kondo effect decreases
exponentially with decreasing $t^{\prime }$ and therefore for small enough $%
t^{\prime }$ the Kondo effect is destroyed by a Zeeman term or temperature
small in comparison with the other energy scales in the problem. Here, we
assume this situation. The effect of a Zeeman term on the conductance in the
Kondo regime, has been explicitly shown mapping the problem to an effective
Anderson model and solving it by non-perturbative methods.~\cite{ihm} In
addition, the main conclusions of this work are of qualitative nature and
are not affected by the accuracy of Eq.~(\ref{tra}).

The conductance is $G=(ne^{2}/h)T(\mu ,V_{g},\phi )$, where $n=1$ or 2
depending if the spin degeneracy is broken or not,~\cite{ihm} and $\mu $ is
the Fermi level, which we set as zero (half-filled leads). As the gate
voltage $V_{g}$ is varied a peak in the conductance is obtained when there
is a degeneracy in the ground state of the ring for two consecutive number
of particles: $E_{g}(N+1)=E_{g}(N)$, where $E_{g}(N)$ is the ground state
energy of $H_{\mathrm{ring}}$ with $N$ electrons. To simplify the
discussion, and without loss of generality, we assume that we start with $%
N+1 $ electrons in the ring and apply a negative gate voltage in such a way
that a peak in the conductance is obtained at a critical value $V_{g}^{c}$
when the number of electrons in the ring changes from $N+1$ to $N$
electrons. Note that Eq.~(\ref{tra}) formally gives more peaks in the
transmittance when at several values of $V_{g}<V_{g}^{c}$ excited states $%
|e\rangle $ of $N$ electrons are reached ($E_{e}(N)=E_{g}(N+1)$). However
these peaks are in principle not experimentally accessible since the ground
state has less than $N+1$ particles for $V_{g}<V_{g}^{c}$. In any case, the
information on the excited states of $N$ electrons is relevant to explain
not only the position of the dips as a function of flux $\phi $ in the
integrated transmittance,~\cite{Jagla,meden,nos1} but also the values of the
flux at which particular reductions of the intensity of the observable first
peak in the conductance are obtained.

\subsection{Dips in the conductance}

The values of the flux $\phi _{d}$ for which dips or reduced conductances
are expected, correspond to some particular crossings of the energy levels
of $N$ electrons. Far from these crossings, and to leading order in $%
t^{\prime }$, the transmittance as a function of gate voltage $V_{g}$ has a
Lorentzian shape reaching the maximum value ($T=1$) for $V_{g}$ such that $%
E_{e}(N)=E_{g}(N+1)$ (where the subscript $e$ refers to any state in the
subspace of $N$ electrons) and half-width at half maximum 
\begin{equation}
w_{e}=\frac{2(t^{\prime })^{2}}{t}\left\vert \langle e|c_{0\sigma }|g\rangle
\right\vert ^{2},  \label{w}
\end{equation}%
where $|g\rangle $ is the ground state in the subspace of $N+1$ electrons
assumed non-degenerate.~\cite{ihm} This assumption is always true in the
presence of a small Zeeman magnetic field except at particular values of the
flux that do not correspond to $\phi _{d}$, which are not relevant for the
present analysis.

Keeping this assumption, some conclusions can be drawn for the general case
using symmetry arguments. Using the Lehman's representation, the part of the
Green's function $G_{0j}^{\mathrm{R}}(\omega )$ that enters the
transmittance [Eq.~(\ref{et})] when a particle is destroyed is: 
\begin{equation}
G_{0j}^{\mathrm{R}}(\omega )=\sum_{e}\frac{\langle g|c_{j\sigma }^{\dagger
}|e\rangle \langle e|c_{0\sigma }|g\rangle }{\omega +E_{e}-E_{g}}.
\label{g1}
\end{equation}

Since the ring is invariant under rotations $R$ that map each site to its
consecutive one $Rc_{j\sigma }^{\dagger }R^{\dagger }=c_{j+1\sigma
}^{\dagger }$, the eigenstates of $H_{\mathrm{ring}}$ are also eigenstates
of $R$. Denoting as $K_{\nu }$ the wave vector of the state $|\nu \rangle $,
one has $R|\nu \rangle =\exp (iK_{\nu })|\nu \rangle $, and therefore $%
\langle g|c_{j\sigma }^{\dagger }|e\rangle =$ $\exp [ij(K_{g}-K_{e})]\langle
g|c_{0\sigma }^{\dagger }|e\rangle $. Replacing above one obtains 
\begin{equation}
G_{0j}^{\mathrm{R}}(\omega )=\sum_{e}\frac{e^{-ij(K_{e}-K_{g})}|\langle
e|c_{0\sigma }|g\rangle |^{2}}{\omega +E_{e}-E_{g}}.  \label{gr}
\end{equation}

At the values of the flux for which two states of $N$ electrons $|e\rangle $
and $|e^{\prime }\rangle $ are degenerate, assuming that the corresponding
matrix elements entering Eq.~(\ref{gr}) are nonzero, clearly only these two
states  (which correspond to the dominant poles of $G_{0j}^{%
\mathrm{R}}(\omega )$) contribute significantly to the Green's function at
the Fermi energy ($\omega =\mu =0$) when $V_{g}$ (which displaces all $E_{e}$
rigidly with respect to $E_{g}$) is tuned in such a way that $%
E_{e}=E_{e^{\prime }}\sim E_{g}$. Replacing Eq.~(\ref{gr}) with only these
two leading terms in Eqs.~(\ref{tra}) and (\ref{et}) one has an analytical
expression for $T(\mu ,V_{g},\phi )$ in terms of two matrix elements,
proportional to $w_{e}$ and $w_{e^{\prime }}$ (Eq.~(\ref{w})) and a relative
phase $\beta =\exp [iM(K_{e^{\prime }}-K_{e})]$ between them in $G_{0M}^{%
\mathrm{R}}(\omega )$ (Eq.~(\ref{gr})). It is easy to see that if $\beta =1$%
, the transmittance for $E_{e}=E_{e^{\prime }}$ has the same form for a
non-degenerate state, but with a width $w_{e}+w_{e^{\prime }}$ equal to the
sum of the individual ones. In this case, nothing dramatic happens. In
particular the integrated transmittance in a window which includes those
levels as a function of flux does not change at the crossing between $E_{e}$
and $E_{e^{\prime }}$.

\begin{figure}[tbp]
\includegraphics[width=7.8cm]{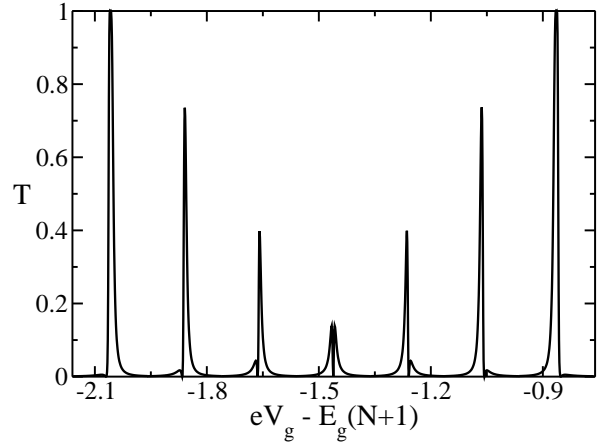}
\caption{Low energy part of the transmittance as a function of gate voltage
for a half-filled ring of $L=6$ sites, $J=0.03t$, $t^{\prime}=0.3t$, and $M=3
$ near a level crossing of the intermediate states with $N=5$ particles. The
crossing is at $\protect\phi _{d}\simeq 0.07743 \protect\pi$ and the
different curves from left to right correspond to $(\protect\phi - \protect%
\phi _{d}) /\protect\pi= -0.04$, $-0.02$, $-0.01$, $0$, $0.01$, $0.02$, and $%
0.04$.}
\label{dfpeaks}
\end{figure}

If, however, $\beta \neq 1$, $G_{0M}^{\mathrm{R}}(\omega )$ and thus the
transmittance, which is proportional to $|G_{0M}^{\mathrm{R}}(\omega )|^{2}$%
, (Eqs.~(\ref{tra}) and (\ref{et})), are reduced near the crossing. This
effect is more noticeable if $\beta =-1$ and $w_{e}=w_{e^{\prime }}$. In any
case, if $\beta \neq 1$, the above mentioned analytical expression \emph{%
vanishes} at the crossing point and the transmittance as a function of gate
voltage shows a peak with a dip inside (Fig.~\ref{dfpeaks}).

 Note that these results are independent of the particular model
used. In particular if the conductance through a ring with $%
N^{\prime }=N+1$ electrons is measured, with leads connected near 180
degrees ($M=L/2$) and the ground state for $N^{\prime }\pm 1$ electrons has
a level crossing involving two wave vectors differing in $2n\pi /L$ with $n$
odd for a given flux, then the conductance shows a dip at that flux.

\section{Analytical results for $\bm{J=0}$}

The general results described above can be made more explicit for $J=0$. In
this case, the model is equivalent to the Hubbard model with infinite
on-site repulsion $U$, for which the wave function can be factorized into a
spin and a charge part.~\cite{ogata,caspers,nos1} Therefore, spin-charge
separation becomes apparent. For each spin state, the system can be mapped
into a spinless model with an effective flux which depends on the total
spin. While the analysis given below can be made using the Bethe ansatz
formalism,~\cite{ogata} here we follow the elegant method of Caspers and
Ilske.~\cite{caspers} For a system of $N$ particles one can construct
spin-wave functions which transform under the irreducible representations of
the group $C_{N}$ of cyclic permutations of the $N$ spins of the $L$-site
system. Each of these representations is labeled by a ``spin'' wave vector $%
k_{s}=2\pi n_{s}/N$, where the integer $n_{s}$ characterizes the spin wave
function. The total energy and momentum (in an appropriate gauge) of any
state of the ring have simple expressions: 
\begin{eqnarray}
E &=&-2t\sum_{l=1}^{N}\cos (k_{l}),\;\;k_{l}=\frac{2\pi n_{l} +\phi _{%
\mathrm{eff}}}{L},  \label{ene} \\
K &=&\sum k_{l}=\left[ 2\pi (n_{c}+n_{s})+N\phi \right] /L,  \label{k} \\
\phi _{\mathrm{eff}}&=&\phi +k_{s}=\phi +\frac{2\pi }{N}n_{s},
\label{phieff}
\end{eqnarray}
where the integers $n_{l}$ characterize the charge part of the wave function
and $n_{c}=\sum n_{l}$.

The calculation of the Green's functions becomes very involved due to
difficulties in handling the wave functions. However, even without
calculating the matrix elements entering Eq.~(\ref{gr}), we can predict the
positions of dips in the transmittance from a knowledge of the energies and
momenta of the eigenstates.

\begin{figure}[tbp]
\includegraphics[width=8cm]{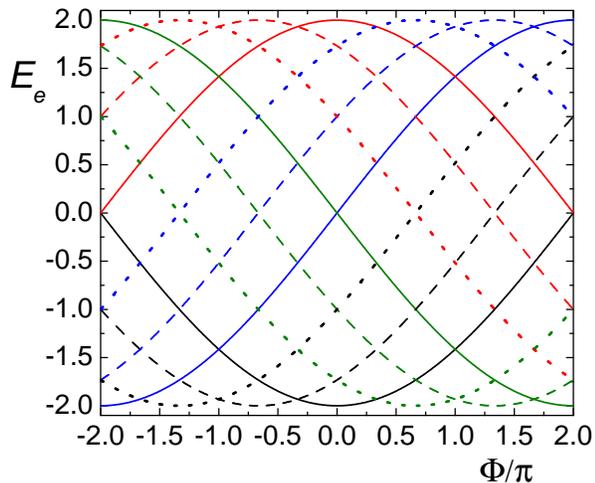}
\caption{(Color online) Energy levels for a system of four sites and three
electrons for $J=0$. Solid lines correspond to the spin quantum number $n_s=0
$, dashed lines to $n_s=-2$ and dotted lines to $n_s=-1$. The different
colors indicate different charge configurations.}
\label{df1}
\end{figure}

Let us analyze first the level crossings for $L$ even and $M=L/2$ (the most
studied case~\cite{Jagla,meden,nos1}) with $N$ odd. We recall the reader
that we start with a system with $N+1$ electrons and create a hole, leading
to an intermediate state with $N$ electrons. The energy levels for the case
with $L=4$, $N=3$ are shown in Fig.~\ref{df1}. In the subspace with $N$
electrons, the ground state $|0\rangle $ (with energy $E_{0}(\phi )$) for $%
n_{s}=\phi =0$ has occupied momenta $k_{l}=2\pi n_{l}/L$, with consecutive $%
n_{l}=-(N-1)/2$, $1-(N-1)/2$, \dots, $(N-1)/2$, and therefore $n_{c}=0$. As
the flux increases to $2\pi $, this state evolves to the state $|1\rangle $,
in which $n_{l}=-(N-1)/2$ is replaced by $(N+1)/2$ with $n_{c}=N$. Therefore
we can write $E_{1}(0)=E_{0}(2\pi )$. The lower ``charge band'' (used
usually to integrate the transmittance~\cite{nos1}) extends between $E_{0}(0)
$ and $E_{0}(2\pi )$. Clearly from Eq.~(\ref{ene}) $E_{0}(-\phi )=E_{0}(\phi )
$ and a state $|-1\rangle $ exists with all $k_{l}$ opposite to those of $%
|1\rangle $. In the enlarged interval $-2\pi \leq \phi \leq 2\pi $, the
ground state energy is reached for $E_{1}(-2\pi )=E_{0}(0)=E_{-1}(2\pi )$
when $n_{s}=0$ (full lines in Fig.~\ref{df1}).  Now keeping the same charge
quantum numbers as $|0\rangle $, but allowing spin states with $n_{s}\neq 0$%
, new eigenfunctions appear, whose energies and momenta are given by $%
E_{0}^{n_{s}}(\phi )=E_{0}(\phi +2\pi n_{s}/N)$, $K=(2\pi n_{s}+N\phi)/L$.
These states reach the ground state energy at $\phi=-2\pi n_{s}/N$ (dashed
and dotted lines in Fig.~\ref{df1}).~\cite{note} The crossings of energy
levels at low energies take place at intermediate points between any two of
these minima $\phi =-\pi (n_{s}+n_{s}^{\prime })/N$. When $%
n_{s}+n_{s}^{\prime } $ is odd (even) the relative phase $\beta =\exp
[iL(K_{e^{\prime }}-K_{e})/2]=\exp \left[ i(n_{s}^{\prime }-n_{s})\right] $
\ $=-1$ (1) and there is (there is not) a dip in the integrated
transmittance. Therefore, the positions of the dips are located at 
\begin{equation}
\phi _{d}=\pi (2n+1)/N,  \label{dip}
\end{equation}
with $n$ integer. These are also the positions where crossings in the
(experimentally accessible) ground state for $N$ particles take place ($%
n_{s}^{\prime }-n_{s}=\pm 1$, see Fig.~\ref{df1}).~\cite{note}

\begin{figure}[tbp]
\includegraphics[width=8cm]{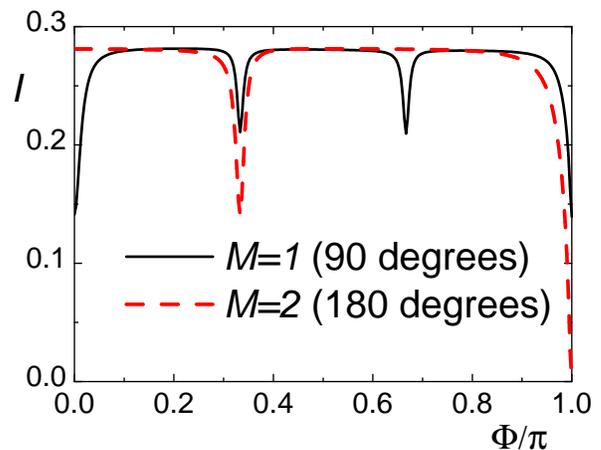}
\caption{(Color online) Total integral of the transmittance for a system of $%
L=4$ sites and $N+1=4$ electrons for $J=0.01$, $t^{\prime}=0.3t$ and two
positions of the leads $M$.}
\label{df2}
\end{figure}

The same expression is valid for $N$ even but in this case for $n_{s}=0$,
the minimum energy lies at $\phi =\pi $, $E_{0}(2\pi -\phi )=E_{0}(\phi )$
and the crossings occur when $\phi =\pi -\pi (n_{s}+n_{s}^{\prime })/N$. 

 Performing the sums in Eqs.~(\ref{ene}) and (\ref{k}) using the
quantum numbers that lead to the minimum energy, as explained above,
analytical expressions are obtained for the ground state energy and momentum
as a function of flux for $L$ sites and $N$ particles, $E_{g}(L,N,\phi )$
and $K_{g}(L,N,\phi )$. Defining

\begin{eqnarray}
\widetilde\phi&=&\phi \quad\qquad\text{for odd }N,  \nonumber \\
\widetilde\phi&=&\phi -\pi \,\quad\text{for even }N,  \label{phip1}
\end{eqnarray}%
and writing $\widetilde\phi$ in the form

\begin{equation}
\widetilde\phi=\text{nint}\left[ \frac{N\widetilde\phi}{2\pi }\right] \frac{2\pi }{N}+\widetilde\phi _{r},  \label{phip2}
\end{equation}%
where nint$\left[ x\right] $ denotes the nearest integer to $x$, one obtains

\begin{eqnarray}
E_{g}(L,N,\phi ) &=&-2t\frac{\sin (N\pi /L)\cos (\widetilde\phi _{r}/L)}{%
\sin (\pi /L)},  \label{eg} \\
K_{g}(L,N,\phi ) &=&N\widetilde\phi/L.  \label{kg}
\end{eqnarray}

For odd $N$, the transmittance vanishes at $\phi =\pi $ due to the
reflection symmetry of the system.~\cite{nos1} This argument does not work
for even $N$ because the ground state of $H_{\mathrm{ring}}$ has orbital
degeneracy for $N+1$ electrons at $\phi =\pi $. It cannot be applied either
for $M\neq L/2$ (where the reflection symmetry is lost~\cite{ihm}) or if the
model includes hopping at large distances (as in Ref.~\onlinecite{meden}).

If the ring is connected to the leads at a distance $M\neq L/2$, the dips at 
$\phi _{d}$ are less intense, because $\beta \neq -1$ if $%
n_{s}+n_{s}^{\prime }$ is odd. However, also $\beta \neq 1$ if $%
n_{s}+n_{s}^{\prime }$ is even and therefore, new dips appear at the
remaining crossings 
\begin{equation}
\phi _{d}^{\prime }=2\pi n/N,  \label{dipp}
\end{equation}
with $n$ integer.

The same arguments can be repeated for states with other charge quantum
numbers lying at higher energies.

These results have been verified numerically and generalize those obtained
previously for $L$ even, $M=L/2$ and odd $N$.~\cite{nos1} In particular, the
total integrated transmittance for a half-filled ring with $L=4$ (so $%
N=L-1=3)$ is shown in Fig.~\ref{df2} for two positions of the drain lead, at 
$M=1$ (90-degree configuration) and $M=L/2$ (180-degree).

Some particular crossings satisfying Eq.~(\ref{dip}) do not lead to dips
because one of the matrix elements of Eq.~(\ref{gr}) vanishes as a
consequence of selection rules related with the total spin. For example, for
the system of Fig.~\ref{df2} the ground state with 4 electrons is a singlet,
and one of the states with lowest energy for $N=3$ that cross at $\phi
_{d}\neq \pi $ is a spin quartet which cannot be accessed destroying an
electron from a singlet. Therefore, in this system, no depression of the
conductance can be observed at equilibrium at zero temperature (for which
only the ground state for any number of particles is accessible) at fluxes
different from half a flux quantum ($\phi _{d}=\pi $). At this flux a
depression of the conductance is already expected for a non-interacting
system. However, this is not the general case, and with increasing number of
particles, states of low total spin are part of the ground state for any
flux. Already for 5 particles, the 32 spin wave functions can be classified
as one sextuplet with $n_{s}=0$, one quadruplet for each of the four $%
n_{s}\neq 0$ and five doublets, one for each of the non-equivalent $n_{s}$.
Addition of $J$, in general favors the lowest total spin (some exceptions
for low $J$ will be discussed below).

Note that for a half-filled system, the energy is zero for $J=0$ independent
of the spin configuration (see Eq.~(\ref{eg})). Therefore addition of an
antiferromagnetic exchange favors the lowest total spin: 0 (1/2) for an even
(odd) number of particles.

To end this section we note that the spin velocity in the limit $%
J\rightarrow 0$ depends strongly on flux. It can be shown that for large $N$%
, it is $v_{s}=v_{c}/N^{2}$ or $v_{s}=v_{c}/N$ for $\phi _{\mathrm{eff}}$
that leads to the minimum or maximum energy respectively.

\section{Numerical results}

In this section we present numerical results for the transmittance, obtained
diagonalizing the ring using Davidson's method~\cite{david} in order to obtain the
Green's functions, which replaced in Eqs.~(\ref{tra}) and (\ref{et}), give
the transmittance. The systems studied are represented in Fig.~\ref{dfhepta}%
. In contrast to previous work,~\cite{Jagla,meden,nos1,julian} we
concentrate on the first peak in the transmittance as the gate voltage is
decreased, which is experimentally accessible at equilibrium and low
temperatures. For weakly coupled rings, this means that one has a system
with $N+1$ electrons in the ring, a hole enters it from one of the leads,
interacting with low-energy intermediate states with $N$ electrons, before
leaving the ring at the other lead, while the ring returns to the ground
state. Therefore the conductance gives information on the low-energy
eigenstates of the ring and, as shown above for $J\rightarrow 0$, on the
separation of charge and spin degrees of freedom in a strongly interacting
system. Similar results can be obtained for increasing gate voltage. In any
case the electronic structure of the low-energy intermediate states is
reflected.

In Fig.~\ref{dfpeaks} we show the transmittance for a half-filled ring of 6
sites connected with the leads at opposite sites (see Fig.~\ref{dfhepta}
(a)) near a crossing of excited states with $N=5$ particles. The value of $%
M=L/2=3$ and the particular crossing were chosen so that according to the
previous sections, a large negative interference is expected, leading to a
depressed conductance. For $J=0$, the crossing occurs at $\phi _{d}=\pi/5$,
but finite $J$ displaces it to smaller values. We have chosen a finite value
of $J$ to break the degeneracy at $J=0$ between eigenstates with $n_s=0$
with total spin $S=1/2$ and $S=5/2$ for $N=5$, as discussed at the end of
the previous section. In any case, the states with $S=5/2$ do not contribute
to the transmittance since they cannot be reached destroying an electron in
the singlet ground state for 6 electrons.

Near the flux $\phi _{d}$ for which two 5-electron states with $S=1/2$ are
degenerate, both corresponding peaks in the transmittance merge into one,
therefore only one peak is seen for each flux (Fig.~\ref{dfpeaks}). Note
that far from the crossing the width of each peak is given by Eq.~(\ref{w})
and increases with $t^{\prime }$. Near the crossing, two poles of the
Green's functions dominate the transmittance as discussed in Section II C,
and they contribute with opposite sign to it. As a consequence a strong
depression of the conductance takes place at $\phi _{d}$. In particular,
Eqs.~(\ref{tra}) and (\ref{et}) indicate that the transmittance vanishes at $%
\phi =\phi _{d}$ for the value of the gate voltage at which the energies of
both 5-electron states coincide with that of the ground state for 6
electrons. This might be an artifact of these expressions which are
perturbative and are not expected to be valid near this point of triple
degeneracy.~\cite{lobos} However, the physical origin of the depression of
the conductance is clear and should be present in a more elaborate treatment.

\begin{figure}[tbp]
\includegraphics[width=8cm]{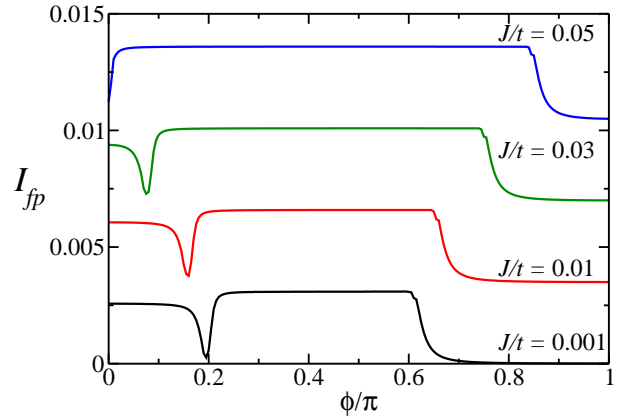}
\caption{(Color online) Intensity of the first peak in the
transmittance as the gate voltage is lowered, $I_{fp}$, as a function of applied
magnetic flux, for a ring of 6 sites and 6 electrons, $M=3$ (Fig.~\protect
\ref{dfhepta} (a)), $t^{\prime}=0.3t$ and several values of $J$. The curves
with $J > 0.001t$ are displaced vertically for the sake of clarity.}
\label{dfL65}
\end{figure}

\begin{figure}[b]
\vspace{0.5cm} \includegraphics[width=8cm]{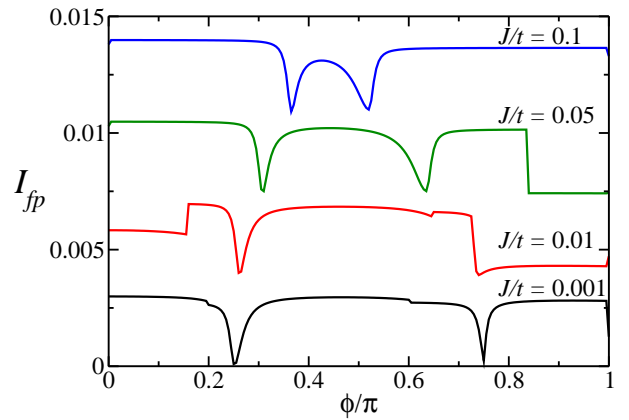}
\caption{(Color online) Same as Fig.~\protect\ref{dfL65} for 5 electrons.}
\label{dfL64}
\end{figure}

To quantify in a systematic way the relative intensity of the conductance
that might be measured in an experimental setup, we integrate the
transmittance given by Eqs.~(\ref{tra}) and (\ref{et}) in a window of gate
voltage $V_{g}$ of width $0.002t$ centered around the degeneracy point
between the ground state for $N+1$ and $N$ electrons. This corresponds to
the {intensity of the first observable peak in the transmittance} as the
gate voltage is lowered. The result as a function of the applied magnetic
flux for the same system of Fig.~\ref{dfpeaks} is represented in Fig.~\ref%
{dfL65}. The curve is symmetric under change of sign of $\phi $ and
therefore we show only the interval $0\leq \phi \leq \pi $. For $%
J\rightarrow 0$ the dips should occur at $\phi _{d}/\pi =0.2$, 0.6 and 1
according to Eq.~(\ref{dip}). However, near 0.6 the ground state for 5
electrons changes from one of total spin $S=1/2$ to $S=3/2$ and the latter
is not accessible destroying an electron in the 6-electron singlet ground
state. Therefore, the transmittance vanishes at the gate voltage for which
the ground state for 5 and 6 electrons have the same energy if $0.6<\phi
\leq \pi $. As a consequence, in the interval shown there is only one dip
present. The position of this dip moves to lower values of $\phi $ with
increasing $J$. As might be expected the region of magnetic flux for which
the ground state with 5 electrons has low total spin ($S=1/2$) increases
with increasing antiferromagnetic exchange $J$.

For a given flux, if the gate voltage is decreased further after the first
peak in the transmittance is observed, the ground state of the ring has now
5 electrons and the next peak corresponds to the degeneracy of the ground
states for 5 and 4 electrons. Fig.~\ref{dfL64} illustrates this situation.
For $J\rightarrow 0$, Eq.~(\ref{dip}) gives dips in the integrated
transmittance for $\phi_d / \pi=0.25$ and 0.75. This agrees with the
numerical calculations. The steps observed for intermediate values of $J$
are again due to jumps in the total spin of one of the ground states (for 4
or 5 electrons). In particular for $\phi / \pi\approx 0.83$ for $J=0.05$, the
total spin of the 5-electron ground state jumps from $S=1/2$ for lower
values of $\phi$ to $S=3/2$ for higher values of the flux.

\begin{figure}[tbp]
\includegraphics[width=8cm]{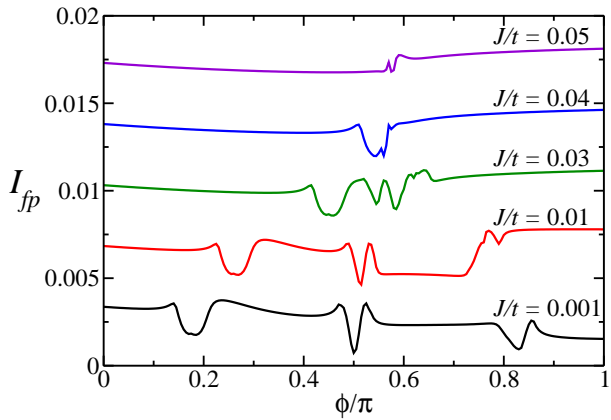}
\caption{(Color online) Same as Fig.~\protect\ref{dfL65} for a ring of 7
sites, 7 electrons and $M=3$ (Fig.~\protect\ref{dfhepta} (b)).}
\label{dfL7}
\end{figure}

Finally in Fig.~\ref{dfL7} we show the results for a half-filled ring of 7
sites with $M=3$ [see Fig.~\ref{dfhepta} (b)]. For low $J$, the dips are
expected at $\phi_d / \pi=1/6$, 1/2 and 5/6 in close agreement with the
numerical results. For increasing $J$ the dips tend to merge into one near $%
\phi=\pi/2$.

Preparing the system with an odd number of electrons, like a half-filled
system with an odd number of sites, has the advantage that more than one
total spin is available for the intermediate states (0 and 1 in this case)
and more dips are observable.

\section{Summary and discussion}

We have shown for the first time that the \emph{equilibrium} conductance
through finite rings described by the $t-J$ model threaded by a magnetic
flux, weakly coupled to conducting leads, at \emph{zero} temperature, shows
depressions at particular values of the flux. In general, for any strongly
correlated model, these depressions are related with level crossings of
excited states and the degree of interference depends on the wave vectors of
these excited states, as described in Section II C.
Previous results involved an integration over an energy window which included several excited 
states and this procedure could cast doubts on the validity of the behaviour for equilibrium 
conductance at zero or very low temperatures.

For the $t-J$ model and $J=0$, using the exactly known energy spectrum of
the isolated ring, we have determined the conditions under which dips in the
integrated transmittance should occur, for different number of particles and
sites of the ring . The position of the dips reflect the particular features
of the spectrum for $J=0$, in which the charge and spin degrees of freedom
are separated at all energies, and not only asymptotically at low energies,
as expected in Luttinger liquids.~\cite{Haldane,Schulz,kv,gia}

In the equilibrium conductance at zero temperature, only the first peak in
the transmittance as a function of gate voltage is accessible in an
experimental setup. Depression of this conductance is expected in general at
certain values of the applied magnetic flux, which are given by Eq.~(\ref%
{dip}) for $J=0$. The negative interference is more marked if the leads are
connected at angles near 180 degrees. These results are confirmed by our
numerical calculations. The positions of depressed conductance are modified
as $J$ is increased in a way which seems difficult to predict. For moderate
values of $J$ it is not clear for us, how to relate any particular position
to a specific change in the spin quantum number. In addition, the number of
these positions seems to decrease with increasing $J$. This is in
qualitative agreement with expectations based on the increase of the spin
velocity,~\cite{Jagla,meden,nos1} which should increase with $J$. In
particular, our results are valid for half-filled systems which we believe
are easier to realize experimentally. Note, however, that due to particular
selection rules explained at the end of Section III, the system size $L$
should be different than four to observe dips in the conductance. In any
case and for any model, if the energy, wave vector and spin of the ground
state is known when one particle is added to or removed from the system, the
position of the dips can be predicted following the arguments of Section II
C.

Due to selection rules related with total spin, more dips are expected for
systems with an odd number of particles.

Our numerical results were based on an expression for the conductance first
used by Jagla and Balseiro,~\cite{Jagla} which is perturbative in the
coupling of the ring with the leads [Eqs.~(\ref{tra}) and (\ref{et})]. This
expression cannot capture non-perturbative effects, like the Kondo physics.~%
\cite{lobos} However, the physics of the depression of the conductance is
present independently of the formalism used to calculate it. Nevertheless a
more elaborate calculation of the conductance would be desirable. One
possibility is to use numerical results for the ring to construct an
effective model for the low-energy physics including the leads, and solve
the model by non-perturbative methods. This approach has been followed in
simpler problems.~\cite{ihm,lobos}

\section*{Acknowledgments}

This investigation was sponsored by PIP 5254 of CONICET and PICT 2006/483 of
the ANPCyT. We are partially supported by CONICET.


\begin{thebibliography}{99}
\bibitem{note} Note that from Eqs.~(\ref{ene}) and (\ref{k}), adding $N$ to $%
n_s$ is equivalent to adding $2 \pi$ to the flux or 1 to all $n_l$ and so we
are allowed to consider values of $n_s$ outside the ``natural" interval $0
\leq n_s \leq N-1$.
\end{thebibliography}
\end{document}